\documentclass[conference]{IEEEtran}
\IEEEoverridecommandlockouts
\usepackage{cite}
\usepackage{amsmath,amssymb,amsfonts}
\usepackage{algorithmic}
\usepackage{graphicx}
\usepackage{textcomp}
\usepackage{xcolor}

\def\BibTeX{{\rm B\kern-.05em{\sc i\kern-.025em b}\kern-.08em
    T\kern-.1667em\lower.7ex\hbox{E}\kern-.125emX}}
\begin{document}

\title{Service-Oriented AoI Modeling and Analysis for Non-Terrestrial Networks\\
\thanks{The work presented in this paper was supported by the National Natural Science Foundation of China under Grand No. 62271168 and the Key R$\&$D Plan of Heilongjiang Province of China under Grand No. JD22A001(Corresponding author: Weixiao Meng).}}
\author{
\IEEEauthorblockN{Zheng Guo$^{\dag}$, Qian Chen$^{\ddag}$, Weixiao Meng$^\dag$}
    \IEEEauthorblockA{$^\dag$ School of Electronic and Information Engineering, Harbin Institute of Technology, Harbin, China}
    \IEEEauthorblockA{$^\ddag$ Department of Electrical and Electronic Engineering, The University of Hong Kong, Hong Kong}
    Email: \tt{zhengguo@stu.hit.edu.cn, qchen@eee.hku.hk, wxmeng@hit.edu.cn}}
\maketitle

\begin{abstract}
To achieve truly seamless global intelligent connectivity, non-terrestrial networks (NTN) mainly composed of low earth orbit (LEO) satellites and drones are recognized as important components of the future 6G network architecture. Meanwhile, the rapid advancement of the Internet of Things (IoT) has led to the proliferation of numerous applications with stringent requirements for timely information delivery. The Age of Information (AoI), a critical performance metric for assessing the freshness of data in information update systems, has gained significant importance in this context. However, existing modeling and analysis work on AoI mainly focuses on terrestrial networks, and the distribution characteristics of ground nodes and the high dynamics of satellites have not been fully considered, which poses challenges for more accurate evaluation. Against this background, we model the ground nodes as a hybrid distribution of Poisson point process (PPP) and Poisson cluster process (PCP) to capture the impact of ground node distribution on the AoI of status update packet transmission supported by UAVs and satellites in NTN, and the visibility and cross-traffic characteristics of satellites are additionally considered. We derived the average AoI for the system in these two different situations and examined the impact of various network parameters on AoI performance.

The simulation results verified the effectiveness of the proposed modeling and analysis method.
\end{abstract}


\section{Introduction}
Despite the rapid evolution of information and communication networks, over 3 billion devices worldwide remain unconnected to the Internet \cite{9722775}. Remote and high-altitude regions, particularly those characterized by challenging terrains like oceans and mountainous areas, confront issues related to inadequate coverage from terrestrial cellular or fiber optic networks due to geographical constraints \cite{9177315}.
Recognized as vital components of the future 6G network architecture, non-terrestrial networks (NTNs) leveraging low-earth orbit (LEO) satellites and unmanned aerial vehicles (UAVs) are poised to overcome the coverage limitations of terrestrial wireless communication, enabling truly global and seamless intelligent connectivity \cite{10355086,10579820}.  

When evaluating system performance, age of information (AoI) is a critical metric for assessing the timeliness and freshness of data \cite{9485125}. It is defined as the disparity between the generation timestamp of the most recent data packet received by the recipient and the present moment \cite{10198349}. Previous studies on AoI predominantly employed various queuing models to characterize systems with varying resource availability and derive average AoI and PAoI \cite{8930830}. The average AoI for M/M/1, M/D/1, and D/M/1 queuing systems under the first-come-first-served (FCFS) strategy was analyzed in \cite{6195689}. Taking into account the number of service servers and queue capacity, \cite{8695040} derived the average AoI with different scheduling strategies. 


However, the majority of existing research has primarily focused on data transmission between ground nodes and terrestrial servers, ignoring the dynamic characteristics and multi-connection characteristics of NTN platforms. Specifically, UAVs can establish connections with multiple ground nodes, and the distribution of ground nodes affects the transmission probability of status update packets. Traditional ground AoI analysis ignores the impact of node distribution and simply models the node distribution as uniform distribution.
Furthermore, satellites can serve not only ground nodes within their coverage area but also act as relay nodes. Differ from the conventional AoI analysis process, it is necessary to focus on the cross-traffic characteristics and visibility of satellite nodes in satellite multi-hop networks. The high dynamics of NTN platforms and the distribution of ground nodes bring challenges to AoI modeling and analysis. Inspired by these facts, this paper investigates the service-oriented AoI modeling and analysis methods in NTN. The main contributions of this paper are outlined as follows

\begin{itemize}
    \item \textbf{Construct a state update system model for NTN}. Ground nodes are represented using a hybrid distribution of PPP-PCP, while the transmission process of state update packets is divided into air-ground and satellite multi-hop network transmissions for further examination.

    \item \textbf{Derive the average AoI for NTN under two queuing systems.}
    We consider two practical queuing systems in UAV and satellite networks: multi-stream M/G/1/1 for UAVs serving multiple ground nodes, and multi-hop serial queuing for satellites with cross-flow dynamics and visibility considerations.

    \item \textbf{Verify the effectiveness of the proposed modeling and analysis method.}
    We extensively discuss the impact of service time distribution variance and the number of satellite hops on the system's AoI.
\end{itemize}
\newpage
\addtolength{\textheight}{-0.729cm} 
The rest of this paper is organized as follows. Section II describes the system model, establishes a node distribution model and a task queuing model. Section III analyzes the AoI of the two proposed data transmission processes. Simulation results are given in Section IV. Finally, the content of this paper is summarized in Section V.

\section{system model}
\begin{figure}[h]
	\centering
	\includegraphics[width = 0.4\textwidth]{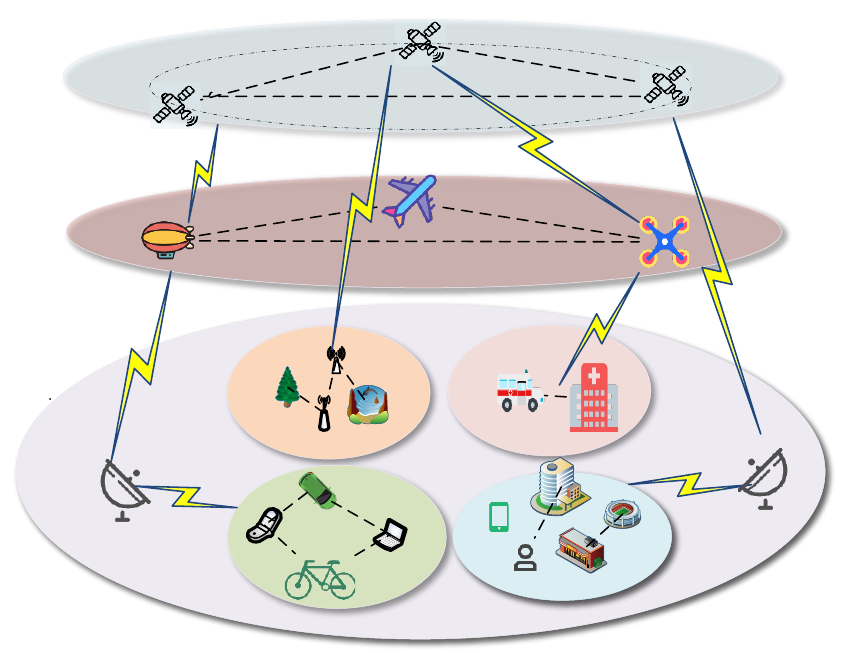}
	\caption{State update network system model, where aerial platforms such as UAVs and satellites provide services to ground nodes. \label{fig:system_model}}
\end{figure}

Considering the communication network shown in Fig. \ref{fig:system_model}, NTN composed of space-based nodes and air-based nodes provides communication services for ground nodes in remote areas. Ground nodes denoted as $U=\left\{u_1, u_2, \ldots, u_n\right\}$. The size of data packets generated by the ground nodes is represented by $S_d$, with the total number of nodes denoted as $N_d$. Air-based nodes include UAVs, aircraft, airships, etc., 
denoted as  $A=\left\{a_1, a_2, \ldots, a_n\right\}$. Space-based nodes constitute satellite networks, primarily consisting of low-orbit satellites, which deliver reliable communication services and extend coverage to areas lacking ground network infrastructure. Let  $K=\left\{k_1, k_2,  \ldots, k_n\right\}$ represent the satellites within the network, and $N_s$ denote the number of visible satellites in the network at any given moment.


\subsection{Node Distribution Model and Wireless Channel Model}
The distribution of UAVs and ground nodes can be characterized using the Poisson point process (PPP) \cite{SAS19-Access}. We represent the positions of UAVs as a homogeneous PPP, denoted by $\Phi_a = {y_i}$, with a density of $\lambda_a$. For the ground nodes, their position denoted as $\Phi_u=\left\{x_i\right\}$ is represented by two distinct point processes. Given the potential for both independent and interrelated services within the network, the ground node positions are modeled as a composite distribution comprising both PPP and PCP \cite{8023870} components.
The proportion of ground nodes adhering to the PPP distribution is denoted by $m_1$ , with a corresponding user density of $\lambda_1$. Similarly, the portion of ground nodes following the PCP distribution is indicated by $m_2$, with central node density of  $\lambda_{p2}$. Within a circular coverage area centered on each central node and with a radius of $r_c$, the ground nodes are uniformly distributed according to independent PPP with a density of $\lambda_{c2}$.
Consequently, the distribution of ground nodes can be defined as
\begin{equation}
\Phi_u=\Phi_{1_x}+\bigcup_{x \in \Phi_p} \Phi_{2_x}.
\end{equation}

Thus, the density of ground nodes in the network can be defined as
\begin{equation}
\lambda_u=m_1 \lambda_1+m_2 \pi r_c^2 \lambda_{p 2} \lambda_{c 2}.
\end{equation}

Furthermore, assuming that a ground node is paired with a UAV offering the maximum average received power, given uniform transmission power across all UAVs, each ground node will naturally access to its nearest UAV.

For wireless transmission channel, assuming the transmission power of the user node is $P_{ui}$, the propagation loss of electromagnetic waves consists of path loss and fading. The path loss at the distance $r$ from the user node can be expressed as $l(r)=r^{-\alpha}$, where $\alpha$ is the path loss index, satisfying $\alpha>2$. The fading model is Rayleigh fading, and the power fading coefficient remains consistent in each time slot and obeys an exponential distribution with a mean value of 1. Assuming that the normalized power of noise is W, both node density $\lambda$ and service arrival rate $\xi$ may affect the relationship between interference and noise.

\subsection{Successful Transmission Probability and Scheduling Probability}

After the SINR threshold of the successful transmission status update packet is defined as $\theta$, the probability that the ground node successfully transmits data packets in slot $j$ is 

\begin{equation}
    p_{s j}=\mathbb{P}\left\{\operatorname{SINR}_j>\theta\right\}.
\end{equation}

Further, the conditional success transmission probability can be obtained as
\begin{equation}
    p_{s j}\left(x_0, \Phi\right)=\mathbb{P}\left[\operatorname{SINR}_j>\theta \mid\left(x_0, \Phi\right)\right].
\end{equation}
where $x_0$ represents the source node that sends status update information, and $\operatorname{SINR}_j$ can be obtained as
\begin{equation}
\operatorname{SINR}_j=\frac{h_{j, x_0} r_0^{-\alpha}}{\sum_{x \in \Phi \backslash\left\{x_0\right\}} h_{j, x}|r|^{-\alpha} \mathbf{1}\left(x \in \Phi\right)+W}. 
\end{equation}


Let $N_j$ denote the number of ground nodes associated with the NTN platform. Assuming the network employs a random scheduling scheme, the scheduling probability of ground nodes following the PPP-PCP mixed distribution is 
 \begin{equation}
     q=\frac{1}{N_j}.
 \end{equation}

\subsection{Task Queuing Model}
The status update packets transmitted in the network can be categorized into two types according to the distance between the source node and the destination node, corresponding to two different task queuing models.

\addtolength{\textheight}{0.729cm}
\subsubsection{The destination node is close to the source node and within the communication coverage area of the low-altitude platform} In this scenario, ground nodes can update status data packets via UAVs without relying on satellite network transmission. 

We model the scenario where a UAV serves multiple ground nodes as a multi-stream M/G/1/1 queuing system. 
IoT packets arrive according to a Poisson distribution, characterized by an arrival rate of $\xi_i$, while the service time follows a general distribution. To maintain stability within the queuing system, the service rate $\rho=\xi/\mu$ must satisfy $\rho<1$. Additionally, due to limited queue capacity, a new packet will be discarded if the previous one is still being serviced when it arrives.

\subsubsection{The destination node is far away from the source node, and the source node is not covered by the UAV network}
Due to the absence of coverage from UAV networks, ground nodes rely on low-earth orbit satellites for transmitting status update data packets. The primary application scenarios encompass IoT deployments for ship and cargo tracking on a global scale. 

The dynamic nature of satellite network topology makes satellites not always visible to ground nodes. Therefore, the visibility probability of satellite nodes represented by $p_a$ is introduced to characterize the service capacity of satellite network queuing systems. 


\subsection{AoI Evolution Model}
\begin{figure}[h]
	\centering
	\includegraphics[width = 0.4\textwidth]{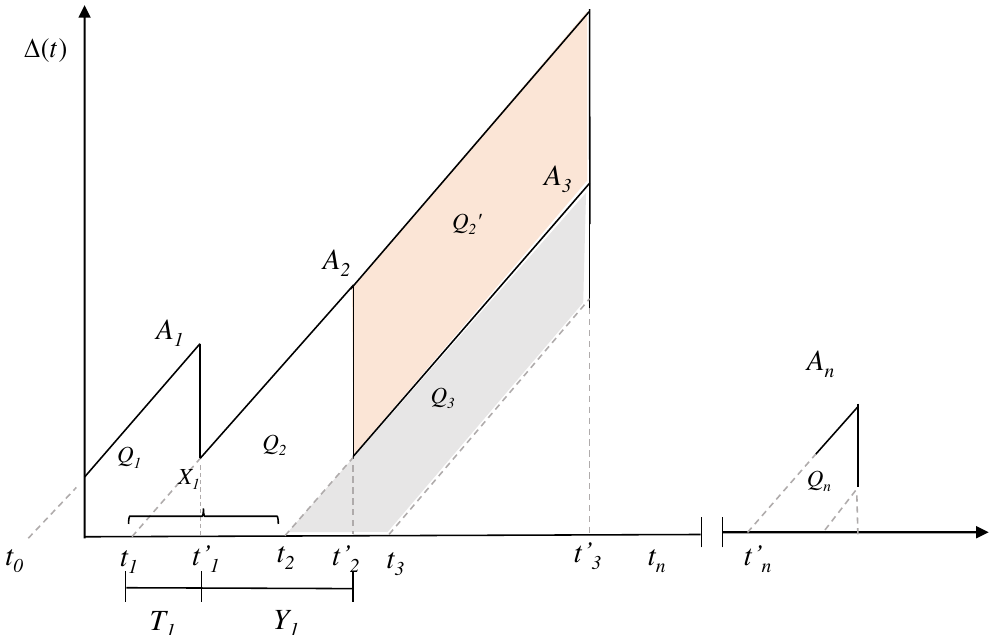}
	\caption{AoI evolution model considering transmission success probability. \label{fig:AOI_model}}
\end{figure}

The AoI $\Delta_t$ is defined as the time it takes for the entire process from the source node to generate a packet to the destination node receiving the latest packet

\begin{equation}
    \Delta_t=t-G(t_k),
\end{equation}
where $G(t_k)$ denotes the time when the state update packet is generated, and $t$ indicates the current time. Fig. \ref{fig:AOI_model} illustrates the variation of instantaneous AoI in the system, where a transmission error occurred in the state update packet that arrived at time $t_2$, and the corresponding instantaneous AoI increased linearly until the packet that arrived at time $t_3$ was successfully received.

For any state update packet in the network, define its arrival time as $t_k$ and departure time as $t_{k'}$. The system service time  $T_k$, arrival time interval  $X_k$, and departure time interval  $Y_k$ can be respectively represented as $T_k=t_{k^{\prime}}-t_k$, $X_k=t_{k}-t_{k-1}$, and $Y_k=t_{k^{\prime}}-t_{{k-1}^{\prime}}$.

The total AoI of the status update system is equal to the area surrounded by the sawtooth line in Fig. \ref{fig:AOI_model}, and the average AoI of the system can be expressed as
\begin{equation}
   \bar{\Delta}=\lim _{T \rightarrow \infty} \frac{1}{T} \int_0^T \Delta(t) d t.
\end{equation}

Then, the relationship between the average AoI and the average peak AoI with  $T_k$ and $Y_k$ can be obtained as follows, where we omitted the subscript $k$

\begin{equation}
\bar{\Delta}=\lim _{T \rightarrow \infty} \frac{N(T)}{T}\left(\frac{\mathrm{E}\left[Y^2\right]}{2}+\mathrm{E}[Y T]\right),\label{eq:AoI}
\end{equation}

\begin{equation}
\bar{\Delta}_p=\lim _{T \rightarrow \infty} \frac{1}{T} \sum_{i=1}^{N(T)} A_i=\mathrm{E}[Y+T].
\end{equation}

\section{Service-oriented {AoI} analysis}

This section considers two different data transmission processes in the network. For the air-ground transmission process, it is modeled as a multi stream M/G/1/1 queuing system. For the satellite multi-hop transmission, the cross traffic process of satellite nodes is first analyzed, and the satellite multi-hop network is modeled as a series M/M/1 queuing system. 

\subsection{AoI Analysis for Air-Ground Transmission}\label{AA}
The transmission process of state update packets in the air-ground network is modeled as a multi-stream M/G/1/1 queuing system. For simplicity, we first analyze the average AoI for a multi-stream M/M/1/1 queuing system and derive a closed-form expression. Following a similar approach, we derive the average AoI for the air-ground data transmission process, taking into account the distribution of ground nodes.

Assume that a UAV serves $M$ ground nodes, each with a status update packet generation rate of $\xi_i$. These packets must be transmitted to the receiving node via wireless channels. Consequently, the total generation rate of status update packets in the system can be expressed as:

\begin{equation}
\xi=\sum_{i=1}^M \xi_i.
\end{equation}

Due to the symmetry of each service flow within the system, the average AoI of all flows can be derived by simply analyzing the average AoI of one of the service flows. Consequently, we specifically analyze the AoI of data stream 1. It's important to note that when a new packet arrives, if the previous packet has not been fully serviced, the newly arrived packet will be discarded. The subscript $k$ is used to denote successfully transmitted data packets.

We note that  $\lim_{T \rightarrow \infty} \frac{N(T)}{T} = \mathbb{E}^{-1}(Y)$, and by substituting it into (\ref{eq:AoI}), we can obtain

\begin{equation}
\bar{\Delta}_1=\mathbb{E}(T)+\frac{\mathbb{E}\left(Y^2\right)}{2 \mathbb{E}(Y)}.\label{eq:AoI1}
\end{equation}

\addtolength{\textheight}{-0.542cm}
For M/M/1/1 non-preemptive queues, we have $\mathbb{E}(T) = 1 / \mu$. To derive the expression in (\ref{eq:AoI1}), the next step is to analyze the distribution of the time interval $Y$.

Let $I$ represent the minimum arrival time interval except for traffic flow 1, i.e. $I=\min \left(X^i\right), i=2, \ldots, M$, which follows an exponential distribution of parameter $\xi-\xi_1$.
When the system moves from state $s_0$ to state $s_1$, it means that it begins to service packets in service flow 1. When the system successfully serves the status update data packet, it returns from state $s_1$ to state $s_0$ again. When the system enters state $s_1^{\prime}$, it starts serving packets from other service flows. After the service is completed in time $U$, the system transitions from state $s_1^{\prime}$ to state $s_0^{\prime}$, and waits in an idle state for the next state update data packet to arrive and provide service for it. It can be found that the system has Markov properties and behaves similarly in state $s_0^{\prime}$ and state $s_0$.


After the system successfully serves the data packet from service flow 1, flow 1 enters state $s_0$. The next state of the system may be  $s_1$ or  $s_1^{\prime}$. The probability of state transition from $s_0$ to $s_1$ is

\begin{equation}
q_1=\mathbb{P}\left(X^{(1)}<I\right) \times p_{s j}\left(x_0, \Phi\right)=\xi_1 p_{s j} / \xi.
\end{equation}

The time the system is in state $s_0$ is defined as $A_1$ and obeys the distribution
\begin{equation}
\mathbb{P}\left(A_1>t\right)=\mathbb{P}\left(X^{(1)}>t \mid X^{(1)}<I\right).
\end{equation}
Similarly, the transition probability from state $s_0$ to $s_1^{\prime}$ is

\begin{equation}
q_2=\left(\xi-\xi_1\right)\left(1-p_{s j}\right) / \xi.
\end{equation}

The corresponding time distribution can be represented as
\begin{equation}
\mathbb{P}\left(A_2>t\right)=\mathbb{P}\left(I>t \mid I<Z^{(1)}\right).
\end{equation}

For service flow 1, the time interval between two consecutive departing data packets corresponds to the total time the system takes to transition from state $s_0$ back to state $s_0$. Thus, $Y$ can be expressed as

\begin{equation}
Y=\sum_{k=1}^{c_1} q_1 A_{1, k}+\sum_{k=1}^{c_2} q_2 A_{2, k}+\sum_{k=1}^{c_3}\left(1-q_1-q_2\right) U_k.
\end{equation}
where ($c_1$, $c_2$, $c_3$) represents the occurrences of $A_1$, $A_2$, and $U$, respectively.

Let $\phi_Y(s)$ denote the moment generating function of the departure time interval

\begin{equation}
\begin{split}
\phi_Y(s) & =\mathbb{E}\left(\mathbb{E}\left(e^{s Y} \mid\left(C_1, C_2, C_3\right)=\left(c_1, c_2, c_3\right)\right)\right) \\
& =\sum_{c_1, c_2, c_3}\left[q_1^{c_1} q_2^{c_2} u^{c_3} \mathbb{R}\left(c_1, c_2, c_3\right) \mathbb{E}\left(e^{s A_1}\right)^{c_1} \right.\\
& \quad \left. \times \mathbb{E}\left(e^{s A_2}\right)^{c_2} \mathbb{E}\left(e^{s U}\right)^{c_3}\right].
\end{split}
\end{equation}

Then, by simplifying the above equation, we can obtain
\begin{equation}
\phi_Y(s)=\frac{\mu \xi_1\left(1-p_{s j}\right)}{s^2 p_{s j}-\left(\xi p_{s j}+\mu\right) s+\mu \xi_1}.\label{eq:moment}
\end{equation}

Take the first and second derivatives of (\ref{eq:moment}) separately, and obtain $\mathbb{E}(Y)$ and $\mathbb{E}(Y^2)$ as follows

\begin{equation}
\mathbb{E}(Y)=\frac{\xi\left(1-p_{s j}\right)+\mu}{\mu \xi_1 p_{s j}},\label{eq:ey1}
\end{equation}

\begin{equation}
\mathbb{E}\left(Y^2\right)=\frac{2\left(\xi p_{s j}+\mu\right)^2-2 \mu \xi_1 p_{s j}^2}{\left(1-p_{s j}\right)^2\left(\mu \xi_1\right)^2}.\label{eq:ey21}
\end{equation}



When comparing the multi-stream M/M/1/1 and M/G/1/1 queues, both share the same semi-Markov chain. The key distinction lies in the service time distribution, specifically the differing waiting times in states $s_1$ and $s_1^{\prime}$. In multi-stream M/G/1/1 non-preemptive queues, the variable $U$ follows the same distribution as the service time $S$, which is also the system time $T$. As a result, the corresponding moment generating function can be expressed as:

\begin{equation}
\phi_Y(s)=\frac{p_{s j} \xi_i \mathbb{E}\left(e^{s T}\right)}{\left(\xi-s p_{s j}^2\right)-\left(\xi-\xi_i\right) \mathbb{E}\left(e^{s T}\right) p_{s j}\left(1-p_{s j}\right)}. \label{eq:moment_mg1}
\end{equation}

Taking the first and second derivatives of (\ref{eq:moment_mg1}) yields $\mathbb{E}(Y)$ and $\mathbb{E}(Y^2)$. Based on the original AoI expression, we can derive the AoI of the multi-stream M/G/1/1 queue considering the transmission error case

\begin{equation}
\Delta_i=\frac{\xi \mathbb{E}(T)+1}{\xi_i\left(1-p_{s j}\right) p_{s j}}+\frac{\xi \mathbb{E}\left(T^2\right) p_{s j}^2}{2\left(\xi p_{s j} \mathbb{E}(T)+1\right)},
\end{equation}

\begin{equation}
\Delta_{p i}=\frac{\left(\xi p_{s j}+\xi_i\right) \mathbb{E}(T)+1}{\xi_i p_{s j}},
\end{equation}
where $\mathbb{E}(T)$ and $\mathbb{E}(T^2)$ are the first and second moments of service time, respectively.

\subsection{AoI Analysis for Satellite Multi-hop Transmission}

Consider a multi-hop satellite transmission process which includes $K$ satellite nodes. Each satellite provides data transmission services to ground nodes in the coverage area and acts as a relay node. The process of sending status update packets from a source node to a remote destination node is modeled as a serial M/M/1 queuing system.

Let $\varepsilon_k$ represent the probability of the $k$-th inter satellite link transmission failure between satellite nodes $k$ and $k+1$. Each node in the satellite network receives cross traffic at a rate of $\theta_k$. The proportion of leaving multi hop connections in the cross traffic entering node $k$ is represented by $\psi_k$. The cross traffic at node $k$ can be represented as

\begin{equation}
\bar{\theta}_k=\sum_{j=1}^k \theta_j \prod_{i=j}^{k-1}\left(1-\psi_i\right)\left(1-\varepsilon_i\right).
\end{equation}

Due to the high-speed mobility of low-orbit satellites, for satellite multi-hop serial queuing systems, two conditions need to be met for successful transmission of status update information:

(1) The signal-to-noise ratio (SINR) at the satellite node is greater than the predetermined threshold, assuming a probability of $P_s$.

(2) The satellite is visible to ground nodes, assuming the visibility probability of the satellite node is $P_a$.

Let $P_r$ represent the probability of successful transmission of state update packets in the series system state. Since conditions (1) and (2) are independent, the transmission success probability of this queuing system model is $P_r=P_s \cdot P_a$. Correspondingly, the probability of transmission failure is $P_c=1-P_r$.

For the AoI evolution model shown in Fig. \ref{fig:AOI_model}, the area of the isosceles trapezoid when packet transmission fails is

\begin{equation}
Q_i^{(1)}=Q_i+Q_{i-1}+Q_{i-1}^{\prime}.\label{eq:Q}
\end{equation}

Furthermore, (\ref{eq:Q}) can be represented by random variables $T$ and $Y$ as follows

\begin{equation}
Q_i^{(1)}=Y_i T_i+Y_{i-1} T_i+\frac{Y_i^2}{2}+\frac{Y_{i-1}^2}{2}.
\end{equation}

Assuming that the occurrence of errors during data transmission is independent, and the probability of correctly transmitting data packets through the first $j$ links is $p_{\mathrm{r}}(j)=\prod_{i=1}^j\left(1-\varepsilon_i\right) p_a$, the average AoI is
\vspace{-5pt}
\begin{equation}
\bar{\Delta}=\xi \sum_{n=0}^{n^{\prime} \rightarrow \infty} p_r(K)^{n^{\prime}-n}\left(1-p_r(K)\right)^n \mathbb{E}\left[Q_i^{(n)}\right],
\end{equation}
which can be represented by $T$ and $Y$ as follows
\vspace{-5pt}
\begin{equation}
\bar{\Delta}_p=\xi \sum_{n=0}^{n^{\prime} \rightarrow \infty} n \cdot p_r(K)^{n^{\prime}-n}\left(1-p_r(K)\right)^n\left(T_i+Y_{i-1}+Y_i\right).
\end{equation}

The total service arrival rate of satellite node $j$ is determined by the successful transmission of services and cross traffic from the source node, which can be expressed as

\begin{equation}
\xi_j=p_r(j) \xi+\bar{\theta}_j.
\end{equation}

Then, the response rate at node $j$ is defined as:

\begin{equation}
\alpha_j=\mu_j-\left(p_r(j) \xi+\bar{\theta}_j\right).
\end{equation}

When all satellites are based on the FCFS service policy, the average AoI of the multi-hop serial M/M/1 queuing system considering transmission errors is

\begin{equation}
\begin{aligned}
\bar{\Delta} & =\xi\left(\begin{array}{l}
\mathbb{E}\left[Y_i T_i\right]+\frac{1-p_r(K)}{p_r(K)} \mathbb{E}\left[Y_{i-1}\right] \mathbb{E}\left[T_i\right] \\
+\frac{1}{2 p_r(K)} \mathbb{E}\left[Y_i^2\right]+\left(\frac{1-p_r(K)}{p_r(K)}\right)^2 \mathbb{E}\left[Y_i\right]^2
\end{array}\right) \\
& =\xi\left(\mathbb{E}\left[Y_i T_i\right]+\sum_{j=1}^N \frac{1-p_s(K)}{p_r(K) \alpha_j^{n_j} \xi} \right. \\
& \quad \left. +\frac{1}{\xi^2 p_r(K)}+\left(\frac{1-p_r(K)}{\xi p_r(K)}\right)^2\right).
\end{aligned} \label{eq:satellite_AoI}
\end{equation}

$\mathbb{E}\left\lceil Y_i T_i\right\rceil$ in (\ref{eq:satellite_AoI}) has a complex coupling relationship between $Y$ and $T$, which is difficult to solve. A simple method is to directly assume that the departure time interval and waiting time of the status update packet are independent, and then obtain an approximation of the average AoI

\vspace{-5pt}
\begin{equation}
\bar{\Delta}=\sum_{j=1}^N \frac{1}{p_r(K) \alpha_j^{n_j}}+\frac{1}{\xi p_r(K)}+\frac{\left(1-p_r(K)\right)^2}{\xi p_r(K)^2}.
\end{equation}

According to $\mathbb{E}\left[W_i Y_i\right] \leq \mathbb{E}\left[T_{i-1}\right] \mathbb{E}\left[Y_i\right] \leq \sum_{j=1}^N \frac{1}{\xi \alpha_j^{n_j}}$, the following upper bound can be obtained for (\ref{eq:satellite_AoI}) 
\vspace{-3pt}
\begin{equation}
\bar{\Delta}_{\text {upper }}=\xi\left(\begin{array}{l}
\sum_{j=1}^N \frac{1}{\xi} \frac{\mu_j^{n_j}+\alpha_j^{n_j}}{\left(\mu_j^{n_j} \alpha_j^{n_j}\right)}+\sum_{j=1}^N \frac{1-p_s(K)}{p_r(K) \alpha_j^{n_j} \xi} \\
+\frac{1}{\xi^2 p_r(K)}+\left(\frac{1-p_r(K)}{\xi p_r(K)}\right)^2.
\end{array}\right).
\end{equation}

\section{simulation results}


In this section, we perform simulation experiments to validate the effectiveness of the proposed modeling and analysis methods. Assume that the ground nodes obey the PPP and PCP distributions in a ratio of 1:1, with $r_c = 5$ m, $\lambda_c = 0.015$, and a total arrival rate of service flows of $x = 3$. For the proposed air-ground UAV multi-stream M/G/1/1 service system, the service time is assumed to follow a gamma distribution, with the variance of the service time distribution denoted by $c^2$.

\begin{figure}[t]
	\centering
 \setlength{\abovecaptionskip}{-5pt}
	\includegraphics[width = 0.35\textwidth]{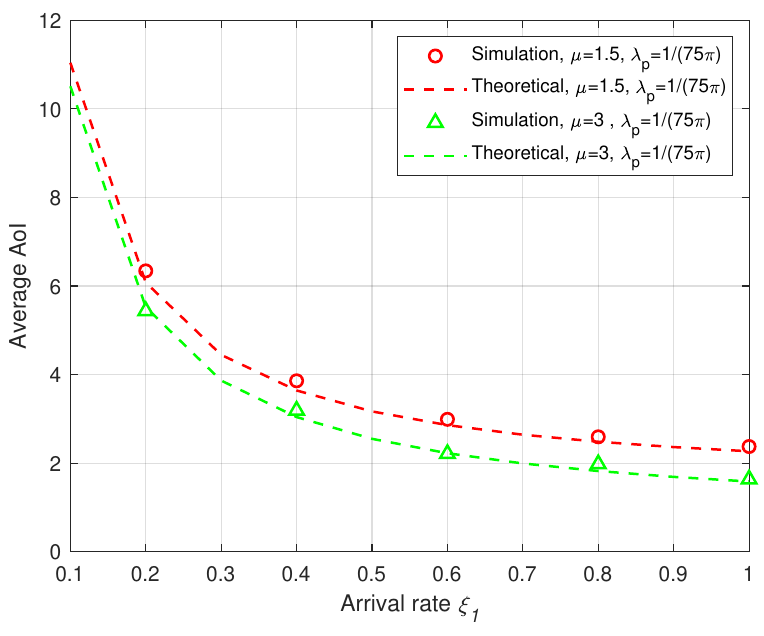}
	\caption{The average AoI of the multi-stream M/M/1/1 queuing system varies with the ground node density and task arrival rate. 
 \vspace{-15pt} 
 \label{fig:simulation1}}
\end{figure}

First, the average AoI of the multi-stream M/M/1/1 queuing system with different service rates is shown in Fig. \ref{fig:simulation1}. The simulation results show strong agreement with the theoretical predictions. With the total service arrival rate remaining constant, the average AoI of service flow 1 steadily declines as the arrival rate increases. This occurs because a higher arrival rate  of service flow 1 increases the probability of packets being serviced in that flow. Additionally, as the system’s service rate increases, the average AoI continues to decrease.

\begin{figure}[t]
	\centering
 \setlength{\abovecaptionskip}{-5pt}
	\includegraphics[width = 0.35\textwidth]{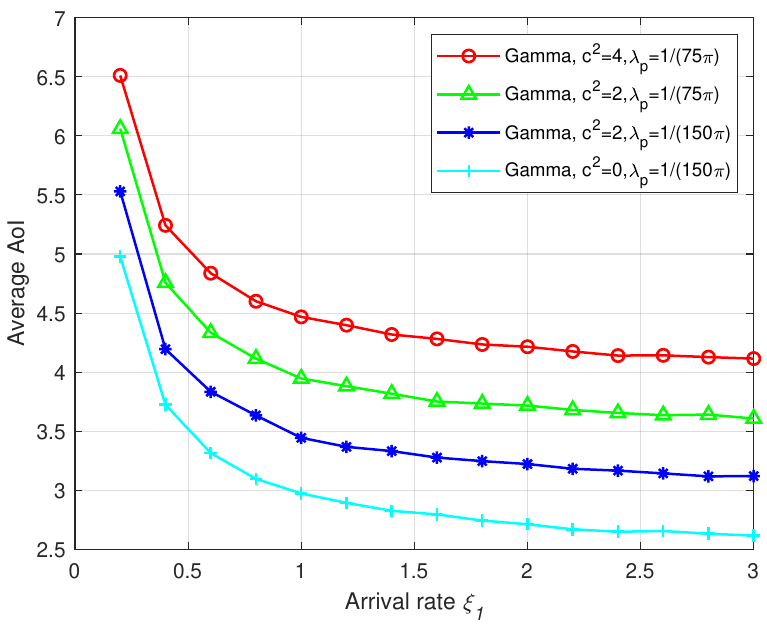}
	\caption{The average AoI of the UAV multi-stream M/G/1/1 queuing system changes with different service time distributions.  
  \label{fig:simulation2}}
  \vspace{-15pt} 
\end{figure}

Fig. \ref{fig:simulation2} depicts the average AoI of our proposed  UAV multi-flow M/G/1/1 queuing system. The system's average AoI is affected by the density of ground nodes and the distribution of service times. A higher variance in the service time distribution leads to an increase in the system's average AoI. Consequently, for a system with constant arrival and service rates, AoI performance deteriorates with significant fluctuations in service times. Additionally, it can be observed that, given the same service time variance, a higher density of ground nodes results in a larger AoI. This suggests that mutual interference between ground nodes negatively impacts the AoI performance of the state update system.

\begin{figure}[t]
	\centering
       \setlength{\abovecaptionskip}{-5pt}
	\includegraphics[width = 0.35\textwidth]{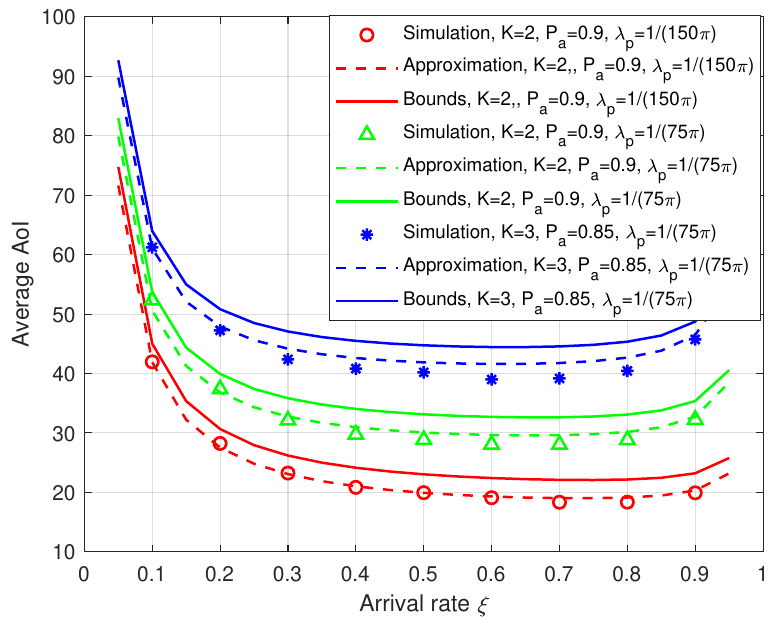}
	\caption{
 The average AoI of the satellite multi-hop serial queuing system varies with the ground node density, task arrival rate, and node sequence number.
  \vspace{-15pt} 
  \label{fig:simulation3}}
\end{figure}

The average AoI of the satellite series queuing system, shown in Fig. \ref{fig:simulation3}, initially decreases and then increases as the arrival rate rises. And higher satellite visibility probability corresponds to lower average AoI. At lower arrival rates, the update frequency of service update data packets is low, resulting in a larger average AoI. However, as the arrival rate escalates, the system load increases, with queuing time becoming the primary factor contributing to the rise in AoI. Additionally, it is observed that when the value of $K$ is small, the approximation effect based on the assumption of independence is better. As the value of $K$ increases, the difference between the approximate curve and the simulation curve widens. This occurs due to the interrelated nature of the data arrival process between nodes in the series queuing system, which is jointly influenced by both the data arrival at the previous node and the satellite coverage area of the current node.


\section{conclusion}

We have proposed a service-oriented AoI modeling and analysis method in NTN. Taking into account the relevance of terrestrial IoT node services, we modeled them as a hybrid distribution of PPP and PCP. Subsequently, we analyzed two different data transmission processes. Specifically, we modeled UAV-ground network transmission as a multi-stream M/G/1/1 queuing system, and then considered the visibility and cross-flow characteristics of satellite nodes to derive the AoI approximation and upper bound for the satellite multi-hop queuing system. Simulation results validate the effectiveness of the proposed method.


\bibliographystyle{IEEEtran}
\bibliography{IEEEabrv,reference}

\end{document}